\documentclass[12pt]{article}
\usepackage{graphicx}
\usepackage{dcolumn}
\usepackage{tensor}
\usepackage{bm}
\usepackage{float}
\usepackage{color}
\usepackage{listings}
\usepackage{slashed}
\usepackage{amsmath}
\usepackage[usenames,dvipsnames]{xcolor}
\usepackage{amssymb}
\usepackage{amsfonts}
\usepackage{enumitem}
\usepackage{tensor}
\usepackage{mathtools}
\usepackage{slashed}
\usepackage{verbatim}
\usepackage{feynmf}
\usepackage{hyperref}
\usepackage{mathrsfs,hyperref,color,float}
\hypersetup{colorlinks=true, linkcolor=blue}
\usepackage{cancel}
\numberwithin{equation}{section}
\usepackage{ulem}
\usepackage{axodraw4j}
\usepackage{pstricks}
\usepackage{color}
\usepackage{cite}

\newcommand{\mO}{\mathcal{O}}
\newcommand{\be}{\begin{equation}}
\newcommand{\ee}{\end{equation}}

\newcommand{\dbar}{\mathchar'26\mkern-12mu d}
\newcommand{\dz}{\mathcal{D}_z^2}
\newcommand{\dzp}{\mathcal{D}_{z'}^2}
\newcommand{\db}{\mathcal{D}_\partial^2}
\newcommand{\dbp}{\mathcal{D}_{\partial'}^2}
\newcommand{\sZ}{\mathcal{Z}}

\newcommand{\binco}[2]{\left( \begin{matrix}{#1}\\{#2}\end{matrix} \right)}

\begin{document}
\title{\textbf{The AdS/CFT partition function, AdS as a lift of a CFT, and holographic RG flow from conformal deformations}} \date{}
\author{Sean Cantrell}
\maketitle
\begin{center}
{\it Department of Physics \& Astronomy, Johns Hopkins University, Baltimore, MD 21218}\\
\noindent\rule{8cm}{0.4pt}
\end{center}
\abstract{Conformal deformations manifest in the AdS/CFT correspondence as boundary conditions on the AdS field.  Heretofore, double-trace deformations have been the primary focus in this context. To better understand multitrace deformations, we revisit the relationship between the generating AdS partition function for a free bulk theory and the boundary CFT partition function subject to arbitrary conformal deformations. The procedure leads us to a formalism that constructs bulk fields from boundary operators.  Using this formalism, we independently replicate the holographic RG flow narrative to go on to interpret the brane used to regulate the AdS theory as a renormalization scale. The scale-dependence of the dilatation spectrum of a boundary theory in the presence of general deformations can be thus understood on the AdS side using this formalism.}
\clearpage

\tableofcontents

\section{Introduction}
The AdS/CFT correspondence has emerged over the last two decades as our best means of understanding quantum gravity.  Originally born as a mapping between states in a string theory living in AdS with compactified extra dimensions and SUSY Yang-Mills theories, the correspondence is now understood broadly as a duality between theories in an AdS background and local conformal theories.  The AdS/CFT correspondence has been understood via two dictionaries:
\begin{enumerate}
\item Taking AdS correlation functions to the boundary and extracting the leading order behavior to recover correlators of dual operators constructed from a local CFT \cite{Susskind:1998dq,Banks:1998dd},
\begin{equation*}
\langle \mO \mO \dots \rangle = \lim_{z\to 0} \langle z^{-\Delta} \phi(z) z^{-\Delta} \phi(z) \dots \rangle
\end{equation*}
\subitem and\\
\item Evaluating the on-shell AdS partition function as a functional of a boundary source, $\phi_b$, and computing CFT correlators in the usual way \cite{Witten:1998qj,Gubser:1998bc},
\begin{equation*}
Z_{\text{CFT}}[\phi_b] = Z_{\text{AdS}}[\phi_b]. \end{equation*}
\end{enumerate}
The dictionaries have been shown to be equivalent in the presence of bulk interactions \cite{Harlow:2011ke}, the intuition being that interactions turn off near the boundary \cite{Giddings:1999qu}, rendering the on-shell description adequate.

The AdS/CFT correspondence also provides a holographic means to understand the renormalization group flow of CFTs as the classical evolution of dual bulk theories in the radial direction of AdS \cite{Erdmenger:2001ja,Sahakian:1999bd,Skenderis:1999mm,Akhmedov:1998vf,Alvarez:1998wr,Porrati:1999ew,Balasubramanian:1999jd,Fukuma:2002sb,Fei:2015oha,Kaplan:2014dia,Komargodski:2011vj,Myers:2010xs,Giombi:2014xxa}.  Explicitly, the radial coordinate is interpreted as the renormalization scale.  Efforts to approach RG flow via entropy-esque quantities and H-theorem type constraints have lead to the a-, F-, and c- theorems \cite{Zamolodchikov:1986gt,Cardy:1988cwa,Komargodski:2011vj,Komargodski:2011xv}, and their associated geometric formulations \cite{Giombi:2014xxa,Fei:2015oha,Myers:2010xs}. 

While few conformal theories appear in nature, deformations can be added to certain ones to more closely reproduce physical theories.  Naturally, we must consider what becomes of the AdS/CFT correspondence when conformality at the boundary is spoiled since, by construction, interactions do not turn off near the boundary.  Conformal deformations were first examined in the context of AdS/CFT in \cite{Aharony:2001pa}, and recent work in conformal dominance \cite{Katz:2013qua,Hogervorst:2014rta} invites their continuing presence. The role of boundary conditions in AdS theories was examined by \cite{Minces:2001zy}, and the connection between boundary conditions and CFT deformations was made explicit by \cite{Witten:2001ua,Berkooz:2002ug}.   

Double-trace deformations have been the predominant focus in the context of AdS/CFT \cite{Witten:2001ua,Hartman:2006dy,Aharony:2001pa,Berkooz:2002ug,Gubser:2002zh,Sarkar:2014jia,Sarkar:2014dma}.  While demonstrative of many salient features of the deformed correspondence, most double-trace techniques, in which the relationship between the bulk and boundary field remains linear while approaching the boundary, do not manifestly apply to more general deformations.  It is also unclear from the literature whether we are instructed to evaluate the AdS partition function on-shell in the usual manner when employing the second dictionary to compute correlators. By this we mean computing the bulk field as the classical solution to the field equations in the presence of a boundary source, $\phi_b$.  Intuition says `no' as this would omit quantum effects from our correlators.

In the first part of this paper, \S \ref{sec:classical} and \S \ref{sec:partition}, we aim to clarify the ambiguities in handling multi-trace deformations and establish a framework that makes manifest the equivalency of the two dictionaries subject to these deformations.  This is achieved by deriving the explicit relationship between the generating bulk partition function and the dual CFT partition function with deforming Lagrangian $W[\mO]$.

This framework then leads to what we call a lift formalism in \S \ref{sec:lift} that is akin to the effort to construct local bulk observables via nonlocal smearing of the boundary \cite{Hamilton:2005ju,Hamilton:2006az,Kabat:2011rz,Xiao:2014uea,Leichenauer:2013kaa,Bena:1999jv}.  The formalism diverges from smearing at the level of operators by providing a nonlinear map between the boundary and the bulk, but similarly achieves the goal of computing bulk correlators from boundary data.  This is done to provide another chapter in the AdS/CFT dictionary and to offer an alternative means of computing Witten diagrams of bulk correlators with boundary deformations, the former seeming particularly useful if boundary data is to be used to understand bulk phenomena.
  
Capitalizing on some results from the lift formalism, we examine in \S \ref{sec:rg_flow} the RG flow triggered by general deformations by interpreting the location of the UV brane used to regulate the AdS theory as a renormalization scale.  Specifically, we compute the scale-dependence of the conformal dimension of the CFT.  We will then briefly comment on potential applications to the recent work of conformal dominance/the truncated Hamiltonian space approach \cite{Yurov:1989yu,Coser:2014lla,Giokas:2011ix,Katz:2013qua,Hogervorst:2014rta}.

We work in the Poincar\'{e} patch of Euclidean $\text{AdS}_{d+1}$ with
\begin{equation}
ds^2 = \frac{1}{z^2} \left( dz^2 + \sum_{i=1}^d dx_i^2 \right)
\end{equation}
and $R_{AdS}=1$.  We consider only CFTs dual to free theories in the bulk. We employ the compact notation $\dbar=\frac{d}{2\pi}$ for integral measures.  We will consider only scalar fields herein, and, where applicable, we will consider a general scaling dimension, $\Delta\ge \frac{d}{2} -1$, of the operator dual to the bulk field; however, with the goal of examining RG flow in mind, we will usually restrict the scaling dimension in the UV to $\Delta = \Delta_- \le \frac{d}{2}$.

We first offer a convenient summary of the results of this paper.

\section{Summary of results}\label{sec:summary}

The generating partition function from which bulk correlators can be computed is
\begin{align}
&Z_{AdS}[J] = \exp \left[  \int d^dx \int dz \,  \frac{1}{2} J(x,z) \phi_{cl}(x,z) \right]  \nonumber \\
&\times \int \mathcal{D} \alpha \mathcal{D} \beta  \exp \bigg[  \int_{z=\epsilon} d^d x \, \left( - \nu \epsilon^{2\nu} \beta^2(x) + W[\alpha(x) + \epsilon^{2\nu} \beta(x) +  \epsilon^{-\Delta_-} \phi_{cl}(x,\epsilon)] \right) \nonumber \\
&+ S_\partial[\alpha] \bigg]. \label{eq:summary_bulk_boundary_partition}
\end{align}
AdS/CFT correlators are generated by functionally differentiating Eq.(\ref{eq:summary_bulk_boundary_partition}) with respect to to the source $J$.  Bulk fields scale to the boundary as $\phi \underset{z \to 0}{\to} \alpha z^{d/2 - \nu} + \beta z^{d/2 + \nu}$, and the classical, particular solution to the field equations is given by $\phi_{cl}(x,z) = \int d^d x' \int_{z'=0}^\infty G(x-x';z,z') J(x',z')$, where $G$ is the bulk-bulk propagator. $W[\mO]$ is the deforming Lagrangian for the dual CFT ($\alpha \to \mO$) with $\Delta_- = \frac{d}{2} -\nu,\, 0<\nu<1$ as the conformal dimension of $\mO$.  From the perspective of the bulk, $W$ is just a boundary term, rendering the bulk theory free. $z=\epsilon \ll 1$ is the location of the UV brane on which the bulk theory terminates that is used to regulate the AdS theory.  $S_\partial$ is a generic conformal action that generates dynamics for $\mO$.  The functional integral over $\alpha$ and the written dependence of $S_\partial$ on $\alpha$ are formalities that simply instruct us to evaluate $\alpha$ as $\mO$ in the CFT.

$\beta$ is an auxiliary field, and integrating it out reproduces Witten's prescription for the boundary conditions:
\begin{align}
\beta = \frac{1}{2\nu} W'[\alpha]. \label{eq:summary_beta_usual}
\end{align}

The partition function given in Eq.(\ref{eq:summary_bulk_boundary_partition}) differs from what usually appears in the literature, where the ``on-shell" action is taken to have the form $S \propto \int_{z=\epsilon} d^d x \, \alpha \beta$ to leading order in $\epsilon$.  In this paper, we argue that these leading terms are canceled and that second order effects must thus be considered to correctly yield Eq.(\ref{eq:summary_beta_usual}) from the on-shell behavior for $\beta$.

It additionally follows from Eq.(\ref{eq:summary_bulk_boundary_partition}) that setting $\epsilon \to 0$, $J \to 0$, and $\phi_b \alpha \subset W[\alpha]$ yields
\begin{align}
Z_{AdS}[\phi_b] = Z_{CFT} [\phi_b],
\end{align}
establishing the second line in Eq.(\ref{eq:summary_bulk_boundary_partition}) as a modified form of the CFT partition function and confirming the second dictionary in the presence of boundary deformations.

Constructing the AdS partition function from the CFT partition function as above is reminiscent of the use of smearing functions to construct AdS operators from their CFT duals.  Smearing functions typically map a tower of operators to a bulk field.  We offer a similar procedure that maps the operators in a different way, but recovers the ultimate goal of constructing AdS correlators from boundary ones.  At the level of operators, we write
\begin{align}
\phi(x,z) =& \int d^d x' L_\alpha (x, x'; z) \alpha(x') + 2\nu \int d^d x' L_\beta (x,x';z) \left( \beta(x') + \beta_0(x') \right), 
\end{align}
where the lift kernels are given by
\begin{align}
L_{\alpha}(x,x';z) =& \int \dbar^d p \, \Gamma(1-\nu) \left( \frac{p}{2} \right)^\nu z^{d/2} I_{-\nu} (p z) e^{i p \cdot (x-x')} \\
L_{\beta}(x,x';z)=& \int \dbar^d p \, \left[ \frac{1}{2} \Gamma(\nu) I_{-\nu} ( p z) - \frac{1}{ \Gamma(1-\nu)}  K_\nu (p z) \right]  \left( \frac{p}{2} \right)^{-\nu} z^{d/2} e^{i p \cdot (x-x')}.
\end{align}
$\beta_0$ acts as a functional derivative when inserted into correlators:
\begin{align}
\beta_0(x) = \frac{1}{2\nu} \frac{\delta}{\delta \alpha(x)},
\end{align}
while $\beta$ takes its usual form of Eq.(\ref{eq:summary_beta_usual}).

These results indeed confirm that the bulk field cannot be computed directly as a classical functional of $\phi_b$ when using the second dictionary in the presence of general deformations.  It must be computed in terms of $\alpha$, which itself can only be computed as a classical functional of $\phi_b$ when computing correlators in the presence of, at most, double-trace deformations.

Using the formalism, we find two interesting results for particular boundary correlators in momentum space:
\begin{align}
\langle W'[\alpha](-p) W'[\alpha](p) +  W''[\alpha](p) \rangle = \frac{\Sigma(p)}{1 - g(p) \Sigma(p)},
\end{align}
and
\begin{align}
\langle \alpha W'[\alpha] \rangle(p) = \frac{g(p) \Sigma(p)}{1 - g(p) \Sigma(p)}. \label{eq:summary_w_ope}
\end{align}
Here, $g$ is the \textit{free} boundary propagator and $\Sigma$ is the sum over two-point function 1PI diagrams at the boundary. Evidently, there is a strong connection between $\beta$ terms in the AdS/CFT dictionary and 1PI diagrams at the boundary. The generally non-vanishing value of the RHS of Eq.(\ref{eq:summary_w_ope}) indicates that the normal ordering one would naively apply to $W'$ when treating it as a multitrace operator is instead applied to the full $\alpha W'$, meaning that $W'$ generally contains both multitrace operators and additional, non-normal ordered operators.  The connection to 1PI diagrams is thus not surprising since $W'$ involves operators that appear in the OPE spectrum of the theory.

We go on to find the conformal dimension of a dual boundary operator depends on RG scale, $\mu$, as
\begin{align}
%
\Delta   \underset{\mu \to \infty}{=}& \left. \frac{ - \mu \partial_\mu \left[ \langle \phi \alpha \rangle (p, \mu) \right]}{ \left[ \langle \phi \alpha \rangle (p, \mu) \right]} \right|_{|p| = \mu} . \label{eq:summary_Delta}
\end{align}
We are instructed to evaluate the correlators with $z = \mu^{-1}$ as a UV brane.  Specifically, we find the bulk-boundary propagator and pull it to the UV brane on which the CFT sits.

It is useful to demonstrate the procedure by computing $\Delta$ for the well-understand example of a double-trace deformation, $\lambda \mO^2$, with UV scaling dimension $\Delta_-$.  We find in momentum space near the boundary
\begin{align}
& \left. \langle \phi\alpha \rangle (p, \mu^{-1}) \right|_{|p| = \mu} \underset{\mu \to \infty}{=} \nonumber \\& -\frac{1}{2} \left[ \frac{1  }{1 + \frac{\Gamma(\nu)}{\Gamma(1-\nu)} \left( \frac{\mu}{2} \right)^{-2\nu} \frac{\lambda}{2}} \right] \frac{\Gamma(\nu)}{\Gamma(1-\nu)} \left( \frac{\mu}{2} \right)^{d/2-\nu}. \label{eq:summary_correlators}
\end{align}
Inserting Eq.(\ref{eq:summary_correlators}) into Eq.(\ref{eq:summary_Delta}) reproduces the known result for double-trace deformations:
\begin{align}
\Delta (\mu) = \Delta_- + \frac{ \lambda }{\mu^{2\nu} + \frac{\lambda}{2\nu}}. \label{eq:summary_double_trace_delta}
\end{align} 
Eq.(\ref{eq:summary_double_trace_delta}) has the simple interpretation that operators in a theory with double-trace deformations with scaling dimension $\Delta_-$ in the UV ($\mu \to \infty$) flow to operators with scaling dimensions $\Delta_+=\Delta_- + 2\nu$ in the IR ($\mu \to 0$).

\section{Boundary conditions and classical bulk fields}\label{sec:classical}

Conformal deformations manifest as boundary conditions in AdS.  Witten originally proposed that a deformation of the form
\begin{equation}
S_W[\mO] = \int d^d x \, W[\mO],
\end{equation}
where $\mO$ is a generalized free field with scaling dimension $\Delta$, should be interpreted as a boundary action in AdS. For dual bulk fields that scale to the boundary as
\begin{equation}
\phi(x,z) \underset{z \to 0}{\to} \alpha(x) z^\Delta + \beta(x) z^{d - \Delta}, \label{eq:phi_scaling}
\end{equation}
the boundary term becomes $S_W[\alpha]$.  This leads to the following constraint on the asymptotic behavior\footnote{The coefficient $\frac{1}{d - 2\Delta}$ is actually dependent on the normalization of $\mO$.  Here, we take $\alpha = \langle \mO \rangle$.}:
\begin{equation}
\beta(x) = \frac{1}{d-2\Delta} \frac{\delta S_W[\alpha]}{\delta \alpha(x)} \label{eq:classical_prescription}.
\end{equation}

This condition was argued to arise classically from an on-shell bulk theory ending on a UV brane at $z=\epsilon \ll 1$ whose action is given by \cite{Aharony:2015afa}
\begin{align}
S[\phi]= S_{bulk}[\phi] + S_{ct}[\phi] + S_W[\phi] =& \int d^d x \int dz \, \sqrt{g} \, \frac{1}{2}\left\{ z^2 ( \partial \phi)^2 - \left[ \left( \frac{d}{2} \right)^2 - \nu^2 \right] \phi^2 \right\} \nonumber \\
&+ \int_{z=\epsilon} d^d x \, \sqrt{g} \mathcal{L}_{ct} + \int_{z=\epsilon} d^d x \, W[ \epsilon^{- \Delta} \phi], \label{eq:deformed_theory}
\end{align}
where
\begin{align}
S_{ct}=\int_{z=\epsilon} d^d x \, \frac{1}{2} \epsilon \Delta \phi^2
\end{align}
is a boundary counter term that ensures the convergence of the on-shell action for $\Delta \ge \frac{d}{2} - 1$ \cite{Klebanov:1999tb,Breitenlohner:1982jf}. For $\Delta \le \frac{d}{2}$, the counter term additionally specifies which of the two viable dual conformal theories sits at the boundary, $\Delta= \Delta_\pm = \frac{d}{2} \pm \nu, 0 \le \nu \le 1$.  The mass-squared has been written in terms of $\nu \equiv \sqrt{\left( \frac{d}{2} \right)^2 - m^2}$ for later convenience.

Classically, Eq.(\ref{eq:deformed_theory}) with $\mathcal{L}_{ct} = \frac{1}{2} \epsilon \Delta \phi^2$ leads to the differential equations
\begin{align}
\left[ \dz + \db \right] \phi =& 0 \label{eq:phidiff}\\
\left[ B_0 + \delta B \right] \phi =&0 \label{eq:phibound},
\end{align}
where
\begin{align}
\dz =& z^{d+1} \partial_z \left[ z^{-d+1} \partial_z \right] + \left[ \left( \frac{d}{2} \right)^2 - \nu^2 \right],\\
\db =&  z^2 \partial_\mu \partial^\mu,\\
B_0 =& \left. \epsilon \partial_z - \Delta \right|_{z=\epsilon} \\
\delta B =& - \left. \epsilon^{d - \Delta} W'[\epsilon^{-\Delta} \phi] \right|_{z= \epsilon}.
\end{align}
Note, the $\mu$ index runs over only the boundary coordinates.  Eqs.(\ref{eq:phibound})\&(\ref{eq:phi_scaling}) then imply, to leading order in $\epsilon$,
\begin{align}
(d - 2\Delta) \beta =& W'[\alpha],
\end{align}
which is, as promised, Eq.(\ref{eq:classical_prescription}).  For now, the classical arguments will be kept to leading order in $\epsilon$; it will be shown later than we must consider second order effects to properly recover the condition in the quantum theory.

We employ the above cumbersome notation to provide a simple, general solution scheme to the classical bulk equations. If we solve the easier problem given by
\begin{align}
\dz \mathcal{Z} =& 0 \label{eq:zdiff}\\
B_0 \sZ =& - \delta B \phi \label{eq:zbound},
\end{align}
and write $\varphi \equiv \phi - \sZ$, we can reduce our task to solving a theory with better known boundary conditions:
\begin{align}
\left[ \dz + \db \right] \varphi =& - \db \sZ\\
B_0  \varphi =&0.
\end{align}
The particular solution is readily apparent:
\begin{align}
\varphi (x, z) =  - \int d^d x' \int_0^\infty dz' \, \sqrt{ g'}\, \dbp G (x - x'; z, z') \sZ(x', z'),
\end{align}
where $G$ is the AdS propagator, modulo factors of $-1$ depending on the convention employed.  For the remainder of this section, we choose $\Delta=\Delta_-$. With this choice, the propagator is 
\begin{align}
G(x; z, z') =& - \int \dbar^d p \int_0^\infty dm \,  (z z')^{\frac{d}{2}} J_{-\nu} (m z) J_{-\nu} (m z') \frac{m}{p^2 + m^2} e^{i p \cdot x}.
\end{align}
The homogenous solution, by construction, must be the same as the $\phi$ solution under the unmodified boundary conditions, and will thus be denoted $\phi_0$.
From this, it follows that
\begin{align}
\phi(x,z) = \phi_0(x,z) + \int d^d x' \int_0^\infty dz' \, \sqrt{g'} \, \dzp G^0 ( x - x'; z, z') \sZ(x', z'). \label{eq:phi_solution}
\end{align}

To verify the procedure, let us consider a boundary source and determine the bulk-boundary propagator. Specifying $W = \epsilon^{d- \Delta + 1} \phi_b \phi $ implies $\delta B = - \epsilon^{d - \Delta} \phi_b$. From Eqs.(\ref{eq:zdiff})\&(\ref{eq:zbound}) we find, to leading order in $\epsilon$, respectively,
\begin{align}
\sZ =& a(x) z^\Delta + b(x) z^{d - \Delta}\\
b(x) =& \frac{1}{2 \nu} \phi_b(x).
\end{align}
The action of the integral kernel in Eq.(\ref{eq:phi_solution}) on $z'^\Delta$ is trivial and, consequently, only $b(x)$ matters. Since we are interested in the particular solution, we discard $\phi_0$, leaving, in momentum space,
\begin{align}
\phi(p,z) =  \left[- \frac{1}{\Gamma(1-\nu)} \left( \frac{p}{2} \right)^{-\nu}  z^{\frac{d}{2}} K_\nu (p z) \right] \phi_b(p).
\end{align}
As expected, the factor in brackets is indeed the momentum space incarnation of the bulk-boundary propagator, 
\begin{align}
\mathcal{K}(x,x';z) =& -\frac{\nu}{\Gamma(1-\nu)} \frac{\Gamma(\Delta_-)}{\pi^{\frac{d}{2}}} \frac{z^{\Delta_-}}{(z^2 + | x - x' |^2 )^{\Delta_-}} .
\end{align}

Next, let us seek the modification to the bulk-boundary propagator with a double trace deformation:
\begin{align}
\mathcal{L}_b = \epsilon^{d - \Delta +1} \phi_b \phi +  \frac{1}{2} \lambda \epsilon^{d - 2 \Delta +1} \phi^2  \implies \delta B[\phi] = - \epsilon^{d-\Delta} \phi_b - \lambda \epsilon^{d - \Delta} \alpha.
\end{align}
This time, Eq.(\ref{eq:zbound}) implies
\begin{align}
b(x) =& \frac{1}{2\nu} \left[ \phi_b(x) + \lambda \alpha(x) \right]\\
a(x)=&0,
\end{align}
and, consequently,
\begin{align}
\phi(p,z) =  -\left[ \frac{1}{\Gamma(1-\nu)} \left( \frac{p}{2} \right)^{-\nu}  z^{\frac{d}{2}} K_\nu (p z) \right] \left[ \phi_b(p) + \lambda \alpha(p) \right]. \label{eq:doubletrace_prelim}
\end{align}
Since we are interested in the particular solution to the boundary source equation, $\alpha$ is not arbitrary and we must solve for it.  Insisting Eq.(\ref{eq:doubletrace_prelim}) match Eq.(\ref{eq:phi_scaling}) and solving for $\alpha$ results in:
\begin{align}
\phi(p,z) =  - \left[ \left(\frac{1}{ 1 + \frac{\Gamma(\nu)}{\Gamma(1-\nu)} \left( \frac{ p}{2} \right)^{-2 \nu} \frac{ \lambda}{2}} \right) \frac{1}{\Gamma(1-\nu)} \left( \frac{p}{2} \right)^{-\nu}  z^{\frac{d}{2}} K_\nu (p z) \right] \phi_b(p).
\end{align}

These results are so far known \cite{Gubser:2002vv,Hartman:2006dy}, but careful attention should be paid to the procedure of throwing out $\phi_0$ and, more precisely, solving for $\alpha$.  It is only with a priori knowledge that the particular solution gives us the bulk-boundary propagator that we have the luxury of demanding $\alpha$ arise only from the source $\phi_b$.  Indeed, had $\phi_0$ not been discarded, it would scale to the boundary as
\begin{align}
\phi_0(x, z) \underset{z \to 0}{\to} z^{- \Delta} A(x),
\end{align}
with $B(x)=0$ ensured by the boundary counter-term.

Restricting to the $\mathcal{K}$ solution has its roots in demanding the bulk field be regular for $z \to \infty$ \cite{Witten:1998qj}, but, since we are permitting the boundary fields to be dynamical, we should reconsider the origin of this restriction.

For general $\alpha$ and $\beta$, the field in the bulk takes the form
\begin{align}
\phi(x,z) =& \int d^d x' L_\alpha (x, x'; z) \alpha(x') + 2\nu \int d^d x' L_\beta (x,x';z) \beta(x'), \label{eq:homogenous_bulk}
\end{align}
where the lift kernels $L_{\alpha,\beta}$ are given by 
\begin{align}
L_{\alpha}(x,x';z) =& \int \dbar^d p \, \Gamma(1-\nu) \left( \frac{p}{2} \right)^\nu z^{d/2} I_{-\nu} (p z) e^{i p \cdot (x-x')} \\
L_{\beta}(x,x';z)=& \int \dbar^d p \, \left[ \frac{1}{2} \Gamma(\nu) I_{-\nu} ( p z) - \frac{1}{ \Gamma(1-\nu)}  K_\nu (p z) \right]  \left( \frac{p}{2} \right)^{-\nu} z^{d/2} e^{i p \cdot (x-x')}.
\end{align}
Demanding $\beta=W'[\alpha]/2\nu$ customarily serve as a source for $\alpha$ requires
\begin{align}
\alpha(x) =& 2\nu \int d^d x' \, g(x - x') W'[\alpha], \label{eq:alpha_constraint}
\end{align}
where the \textit{undeformed} boundary propagator is given by
\begin{align}
g(x - x') =& -\int \dbar^d p \, \frac{\Gamma(\nu)}{2 \Gamma(1-\nu)} \left( \frac{p}{2} \right)^{-2 \nu} e^{i p \cdot (x - x')} \propto \frac{1}{ | \Delta x |^{2 \Delta}}.
\end{align}
Inserting Eq.(\ref{eq:alpha_constraint}) into Eq.(\ref{eq:homogenous_bulk}) yields
\begin{align}
\phi(x,z) = 2 \nu \int d^d x' \, \mathcal{K} (x - x'; z) W'[\alpha],
\end{align}
which is precisely the form achieved by discarding $\phi_0$ in the above formalism.

This indicates that including $S_{ct}$ and $S_W$ in the AdS action cannot be the entire story.  An action for $\alpha$, denoted herein as $S_\partial$, that classically results in Eq.(\ref{eq:alpha_constraint}) must either be generated or appear by explicit insertion.   Given the nature of the constraint, $S_\partial$ should include terms that lead to a dynamical $\alpha$, such as what would be contained in a generalized free theory in the large $N$ limit.  We argue in the following section that $S_\partial$ is generated by integrating out the bulk.

As will be shown in the following section, multi-trace deformations require a proper quantum treatment, and so our chief classical analysis ends here; however, we consider multi-trace deformations and bulk wave functions and demonstrate that using the double-trace techniques of this section requires $\alpha$ to be classical in Appendix \ref{app:classical}.

\section{Bulk and boundary partition functions redux}\label{sec:partition}

Conformal deformations generate interactions and, generally, quantum corrections. Indeed, quartic interactions anomalously break conformal invariance precisely due to the appearance of a renormalization scale arising from loop corrections. Since Witten diagrams will include vertices at the boundary, we should expect the bulk theory to inherit certain quantum effects.  To demonstrate the appearance of such quantum corrections to bulk correlation functions at the level of the partition function, consider a multi-trace deformation constrained to a bulk UV brane at $z=\epsilon$ for a field $\phi$ dual to a conformal operator with scaling dimension $\Delta_-$,
\begin{align}
W[\phi]=& \frac{1}{n} \lambda \epsilon^{- n \Delta} \phi^n (x, \epsilon)\\
\mathcal{L}_{ct} =& \frac{1}{2} \epsilon \Delta_- \phi^2(x,\epsilon).
\end{align}
The generating partition function for bulk correlators is given by\footnote{Here, the source is given its conventional symbol, $J$, which should not be mistaken for the Bessel functions appearing in the propagator.}
\begin{align}
Z[J] = \int \mathcal{D} \phi \, \exp \Bigg[  & \Bigg(  \int d^d x \int dz \, \sqrt{g} \, \frac{1}{2} \left\{  z^2 (\partial \phi)^2 - \left[ \left( \frac{d}{2} \right)^2 - \nu^2 \right] \phi^2  \right\} + J \phi  \nonumber \\
&+ \int_{z=\epsilon} d^d x \, \epsilon^{-d - 1} \left\{ \frac{1}{2}\epsilon \Delta_- \phi^2  + \frac{1}{n} \lambda \epsilon^{d-n\Delta_- + 1} \phi^n \right\} \Bigg) + S_\partial[\phi] \Bigg]. \label{eq:deformed_integral}
\end{align}
We can separate the quantum effects from the classical solutions by changing the integral measure:
\begin{align}
\phi=&\phi_{cl} + \theta\\
J =& \sqrt{g} \left\{ \nabla^2 + \left[ \left( \frac{d}{2} \right)^2 - \nu^2 \right] \right\} \phi_{cl}, \label{eq:classical_field_eqns}
\end{align}
where, additionally, $\phi_{cl}$'s boundary conditions are chosen to minimize the on-shell action,
\begin{align}
\left. \left[ -\epsilon \partial_z \phi_{cl} + \Delta_- \phi_{cl} + \lambda \epsilon^{d- n \Delta_-} \phi_{cl}^{n-1} \right] \right|_{z=\epsilon} =&0, \label{eq:classical_field_bc}
\end{align}
which leads to Eq.(\ref{eq:classical_prescription}). The regular mode of the classical solution, $\alpha$, is additionally forced by $S_\partial$ to reproduce only the particular solution from Eq.(\ref{eq:classical_field_eqns})

Integrating the first term in Eq.(\ref{eq:deformed_integral}) by parts and inserting Eqs.(\ref{eq:classical_field_eqns})\&(\ref{eq:classical_field_bc}) into the result yields
\begin{align}
Z[J] &= \exp \Bigg[   \int d^d x \int dz  \, \frac{1}{2} J \phi_{cl}[J] \Bigg] \nonumber \\
&\times \int \mathcal{D} \theta \, \exp \left[ S[\theta, J=0] \right] \exp \left[   \int_\partial d^d x \,\frac{1}{n} \lambda \epsilon^{ -n\Delta_-}  \sum_{m=2}^{n-1} \binco{n}{m} \theta^m \phi_{cl}^{n-m}[J] + S_\partial( \theta) \right] , \label{eq:deformed_integral2}
\end{align}
where the functional dependence of $\phi_{cl}$ on $J$ has been emphasized. Bulk correlators are specified by the integral kernels in the functional expansion of  Eq.(\ref{eq:deformed_integral2}) in terms of $ J$.  Evidently, the coupling in the second line vanishes for $n\le 2$ and quantum effects are manifestly absent; for $n > 2$, however,  loop effects begin appearing, revealing the short coming of the usual approach taken for double trace deformations. For example, cubic interactions generate loop corrections to the bulk two point function at order $\lambda^2$. Explicitly,
\begin{align}
\langle \phi(x',z') \phi(x,z) \rangle =& \left. \frac{\delta^2}{\delta J(x',z') \delta J(x,z)} Z[J] \right|_{J=0} \nonumber \\
=& \Bigg[ \frac{\delta \phi_{cl}(x,z)}{\delta J(x',z')} + \lambda \epsilon^{-3 \Delta_-} \int d^d y \langle  \,\theta^2(y, \epsilon) \rangle \frac{\delta^2 \phi_{cl}(y,\epsilon)}{\delta J(x',z') \delta J(x,z)} \nonumber \\
&+ \left.  \lambda^2 \epsilon^{- 6 \Delta_-} \int d^d y \int d^d y' \langle \theta^2(y',\epsilon) \theta^2(y,\epsilon) \rangle \frac{\phi_{cl}(y',\epsilon)}{\delta J(x',z')} \frac{ \delta \phi_{cl} (y, \epsilon)}{\delta J(x,z)} \Bigg] \right|_{J=0}.
\end{align}
The factors of $\epsilon^{-3\Delta_-}$ divide out the vanishing parts of the classical bulk correlators as they are taken to the boundary.  The term naively proportional to $\lambda$ vanishes since the vacuum specified by action $S$ does not support vevs of $\theta$ or its composites; nonetheless, the functional derivatives of $\phi_{cl}$ are non-zero and arise from the second harmonic term of $\phi_{cl}$, which itself is proportional to $\lambda$, ensuring the quantum corrections are indeed of order $\lambda^2$. Higher order corrections are implicitly contained in the usual loop term $\langle \theta^2(y') \theta^2 (y) \rangle$; as addressed in Appendix \ref{app:classical}, the bulk-bulk and bulk-boundary propagators do not contribute larger $\lambda$ corrections as they are non-vanishing only for non-vanishing $J$.

It follows that pulling correlators computed according to Eq.(\ref{eq:deformed_integral2}) to the boundary and computing correlators by evaluating the bulk partition function classically after pulling the source to the boundary are inequivalent.  This is not to say the bulk and CFT partition functions are inequivalent, only that we should forgo computing the bulk fields classically.  While the interior of the AdS action may be treated classically in the absence of bulk interactions, quantum effects appear through the boundary conditions.  This suggests that the bulk theory should be understood as a lift of a quantum boundary theory. To clarify, this simply means the bulk is a classical boundary-value problem with quantum dynamics governing the behavior of the boundary conditions.  To formalize this notion, we wish to express the generating bulk partition function as a functional of the boundary fields. We follow the procedure of dividing the generating path integral into IR ($\epsilon+ \delta < z$) and UV ($\epsilon \le z \le \epsilon + \delta$) contributions \cite{Heemskerk:2010hk,Harlow:2011ke},
\begin{align}
Z[J] =& \int_{IR} \mathcal{D} \phi \, \exp \left[ S_{bulk}[\phi]  +  \int d^d x \int_{\epsilon+\delta}^\infty dz \,  J \phi \right]  \nonumber \\
&\times \int_{UV} \mathcal{D} \phi \, \exp  \left[  S[\phi] +  \int d^d x \int_{\epsilon}^{\epsilon+\delta} dz \,  J \phi \right], \label{eq:generating_integral}
\end{align}
and let $\delta \to 0$.  $S$ and $S_{bulk}$ are given by Eq.(\ref{eq:deformed_theory}) and we will choose $\Delta = \Delta_-$ as the scaling dimension of the undeformed boundary theory.   As before, we separate the classical and quantum effects, but break down the quantum corrections in the IR differently:
\begin{align}
\phi(x,z) =& \phi_0(x,z) + \phi_{cl}(x,z) + \theta(x,z), \\
\phi_0(x,z) =& \frac{1}{2\nu} \epsilon^{- \Delta_-} \int d^d x' L_\alpha (x, x'; z) \left[ \Delta_+ \theta(x,\epsilon) - \epsilon \partial_z \theta(x, \epsilon) \right] \nonumber \\
&- \epsilon^{- \Delta_+} \int d^d x' L_\beta (x,x';z)  \left[ \Delta_- \epsilon \theta(x, \epsilon) - \epsilon \partial_z \theta(x,\epsilon) \right], \\
\phi_{cl}(x,z) =& \int d^d x' \int dz' \, G(x -x'; z, z') J(x',z')
\end{align}
The $\phi_0$ term contained in the IR bulk field is, explicitly, a lift of the homogeneous boundary (UV) fields.  As before, $\theta$ is the quantum perturbation. With this ansatz, the generating partition function becomes
\begin{align}
Z[J] \underset{\delta \to 0}{=}&  \exp \left[  \int d^dx \int dz \, \frac{1}{2} J \phi_{cl}  \right]   \int_{IR} \mathcal{D} \theta \exp\left[ S_{bulk}[\theta] \right] \nonumber \\
\nonumber \\
& \times \int \mathcal{D} \theta(\epsilon+ \delta) \mathcal{D} \theta(\epsilon) \bigg[   \int_{z= \epsilon} d^d x \, \left( -  \frac{1}{2} \epsilon^{-d+1} \phi_0 \partial_z \phi_0 +   \frac{1}{2} \epsilon^{-d} \Delta_- \phi_0^2  \right) \nonumber \\
& + S_W[ \phi_{cl} + \phi_0] \bigg].
\end{align}
The path integral variable $\theta(\epsilon+\delta)$ appears only through the derivative $\partial_z \phi_0= \frac{ \theta(\epsilon+\delta) - \theta(\epsilon)}{\delta}$.  It is more convenient to work in terms of the usual functions $\alpha$ and $\beta$ that parameterize the asymptotic behavior of $\phi_0$.  Writing
\begin{align}
\left( \begin{matrix}
\phi_0(\epsilon) \\
\phi_0(\epsilon+\delta)
\end{matrix}
\right)
=
\left( \begin{matrix}
\epsilon^{\Delta_-} & \epsilon^{\Delta_+} \\
(\epsilon+\delta)^{\Delta_-} & (\epsilon+\delta)^{\Delta_+}
\end{matrix}
\right) \left(
\begin{matrix}
\alpha \\
\beta
\end{matrix} \right),
\end{align}
the generating partition function becomes
\begin{align}
Z[J] =& \exp \left[  \int d^dx \int dz \,  \frac{1}{2} J \phi_{cl} \right]  \int_{IR} \mathcal{D} \theta \exp\left[ S_{bulk}[\theta] \right] \nonumber \\
&\times \int \mathcal{D} \alpha \mathcal{D} \beta  \exp\left[  \int d^d x \, \left( - \nu \alpha \beta  - \nu \epsilon^{2\nu} \beta^2 + W[\alpha + \epsilon^{2\nu} \beta +  \epsilon^{-\Delta_-} \phi_{cl}] \right)  \right], \label{eq:bulk_boundary_partition_pre}
\end{align}  
modulo irrelevant factors of $\delta$ and $\epsilon$ that arise from the Jacobian from our change of integral measure.

In the absence of a bulk source and as $\epsilon \to 0$, we expect to recover the correct CFT partition function with the appropriate $S_\partial$.  The derivative terms in the IR partition function may be integrated by parts, leaving only boundary terms, and the bulk fields can be integrated out resulting in
\begin{align}
 &\exp\left[ \frac{1}{2} \int d^d x dz \sqrt{g} \int d^d x' dz' \sqrt{g'} G(x-x';z,z') \frac{\delta}{\delta \theta(x,z)} \frac{\delta}{\delta \theta(x',z')} \right]  \nonumber \\
 &\times \exp\left[ - \int_{z=\epsilon} d^d y \epsilon^{-d+1} \theta \partial_z \theta \right].\label{eq:ir_eval}
\end{align}
We do not offer a rigorous proof that this generates the necessary terms, but point out that for the AdS/CFT dictionary to hold in the absence of deformations and for the counter terms currently used in the literature to be correct, Eq.(\ref{eq:ir_eval}) must evaluate to\footnote{The $S_\partial$ term should be expected from the AdS/CFT story; the $\nu \alpha \beta$ term is necessary to counter the $-\nu \alpha \beta$ term one would obtain by evaluating the bulk action on-shell and integrating by parts. Without canceling it out, the boundary correlators would not be evaluated as expected in the dual CFT.}
\begin{align}
\exp\left[ \int d^d x \left( \nu \alpha \beta + S_\partial[\alpha] \right) \right].
\end{align}
Inserting this into Eq.(\ref{eq:bulk_boundary_partition_pre}) finally yields
\begin{align}
Z_{AdS}[J] =& \exp \left[  \int d^dx \int dz \,  \frac{1}{2} J \phi_{cl} \right]  \nonumber \\
&\times \int \mathcal{D} \alpha \mathcal{D} \beta  \exp\left[  \int d^d x \, \left( - \nu \epsilon^{2\nu} \beta^2 + W[\alpha + \epsilon^{2\nu} \beta +  \epsilon^{-\Delta_-} \phi_{cl}] \right) + S_\partial[\alpha] \right].\label{eq:bulk_boundary_partition}
\end{align}

Evidently, $\beta$ is an auxiliary field.  Integrating it out sets, to leading order in $\epsilon$,
\begin{align}
\beta = \frac{1}{2\nu} W'[\alpha ], \label{eq:beta_redux}
\end{align}
thus recovering the usual boundary conditions.

Meanwhile, the functional integral over $\alpha$ and the written dependence of $S_\partial$ on $\alpha$ are formalities that simply instruct us to evaluate $\alpha$ as $\mO$ (up to normalization factors) given the appropriate boundary CFT.

It is immediately apparent from Eq.(\ref{eq:bulk_boundary_partition}) that
\begin{align}
\lim_{\epsilon \to 0} Z_{AdS}[\phi_b] = Z_{CFT} [\phi_b].
\end{align}

\section{Bulk correlation functions and the lift formalism}

With Eq.(\ref{eq:bulk_boundary_partition}) establishing an appropriate dictionary for multi-trace deformations, we may compute bulk correlators and compare them to expectations from Witten diagrams.

The expression of the bulk partition function in terms of a boundary partition function through the construction of the bulk fields from boundary fields in the previous section encourages the literal interpretation of the bulk theory as a theory of quantum boundary conditions.  This leads us to consider the use of smearing functions to compute bulk correlators from boundary correlators.  We review the use of smearing functions, and go on to develop an alternative, but related, formalism.

\subsection{Correlating} \label{sec:correlators}
It immediately follows from Eq.(\ref{eq:bulk_boundary_partition}) that bulk correlation functions take the form
\begin{align}
\langle \phi(x_1, z_1) \phi(x_2, z_2) \dots  \phi(x_{n-1}, z_{n-1}) \phi(x_n, z_n) \rangle =& G(x_2 - x_1; z_2, z_1)\dots G(x_{n} - x_{n-1}; z_{n}, z_{n-1}) + \{perm\} \nonumber \\
&+ \prod_{i=1}^n \left[ \int  d^d y_i \, \mathcal{K}(x_i - y_i; z_i) \right] \Gamma_n(y_1, y_2, \dots , y_n),
\end{align}
where the vertex function, $\Gamma_n$, arises from derivatives of $W$.  This is in agreement with the form that follows from Witten diagrams.

\begin{figure}[H]
\begin{center}
\includegraphics[scale=1.0]{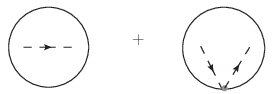}
\caption{\small Witten diagram for the bulk two-point function. The sum over 1PI parts at the boundary is absorbed into the vertex.}\label{fig:bulk_two_point}
\end{center}
\end{figure}

Computed via diagrams, the two-point function, shown in Fig. \ref{fig:bulk_two_point}, for instance, is given in momentum space by
\begin{align}
\langle \phi(-p, z_1) \phi(p, z_2) \rangle =& G(p; z_1, z_2) + \mathcal{K}(p, z_1) \frac{\Sigma(p)}{1 - g(p) \Sigma(p)} \mathcal{K}(p, z_2), \label{eq:two_point_witten}
\end{align}
where $\Sigma(p)$ is the usual sum of 1PI diagrams at the boundary. Using Eq.(\ref{eq:bulk_boundary_partition}), we find, schematically,
\begin{align}
\Gamma_2 =& \langle W'[\alpha] W'[\alpha] +  W''[\alpha] \rangle.
\end{align}
Identifying this with Eq.(\ref{eq:two_point_witten}) requires
\begin{align}
\frac{\Sigma}{1 - g \Sigma} = \langle W'[\alpha] W'[\alpha] +  W''[\alpha] \rangle. \label{eq:1PI_W}
\end{align}

For a double-trace deformation, $W=\frac{1}{2} \lambda \alpha^2$, we find
\begin{align}
\langle W'[\alpha] W'[\alpha] +  W''[\alpha] \rangle =&  \frac{\lambda}{1 - g(p) \lambda},
\end{align}
where the boundary correlators are computed according to the usual CFT rules, including summing over all double trace insertions.  This predicts $\Sigma = \lambda$, as would be expected diagrammatically from adding a mass term. 

\begin{figure}[H]
\begin{center}
\includegraphics[scale=1.0]{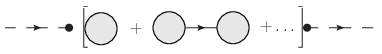}
\caption{\small A diagrammatic representation of the vertex function $\langle \alpha^2 \alpha^2 \rangle$. The 1PI diagrams in brackets are indeed amputated.}\label{fig:cubic_sum}
\end{center}
\end{figure}

From the cubic deformation $W[\alpha] = \frac{1}{6} \lambda \alpha^3$, we find
\begin{align}
\langle W'[\alpha](y_1) W'[\alpha](y_2) =\left( \frac{1}{2} \lambda \right)^2 \langle \alpha^2(y_1) \alpha^2(y_2) \rangle.
\end{align}
The boundary correlator, which is represented diagrammatically as the bracketed factor in Fig. \ref{fig:cubic_sum}, evaluates to\footnote{A factor of 2 appears at each vertex due to symmetry.}
\begin{align}
\left( \frac{1}{2} \lambda \right)^2 \langle \alpha^2(-p) \alpha^2(p) \rangle =& \frac{\Sigma(p)}{1 - g(p) \Sigma(p) },
\end{align}
with the usual cubic 1PI diagram for $\Sigma$,
\begin{align}
\Sigma(p) \propto \lambda^2 \int \dbar^d l \, l^{-2\nu} ( p - l )^{-2\nu} + \dots,
\end{align}
manifestly agreeing with the Witten diagram. 

The bulk three-point function arising from the cubic deformation is similarly easy to assess using Eq.(\ref{eq:bulk_boundary_partition}):
\begin{align}
\langle \phi(x_1,z_1) \phi(x_2,z_2) \phi(x_3,z_3) \rangle =& \prod_{i=1}^3 \left[ \int d^d y_i  \, \mathcal{K}(x_i -y_i; z_1)\right] \nonumber \\
& \times  \Bigg[ \lambda \int d^d y \, \delta^d (y_1 - y) \delta^d (y_2 - y) \delta^d (y_3 - y) \nonumber \\
& + \left( \frac{1}{2} \lambda \right)^3 \langle \alpha^2(y_1) \alpha^2(y_2) \alpha^2(y_3) \rangle \Bigg].
\end{align}
As should be expected, loop effects enter through the boundary correlator $\langle \alpha^2(y_1) \alpha^2(y_2) \alpha^2(y_3) \rangle$.

The agreement between bulk correlators computed via diagrams and those computed using Eq.(\ref{eq:bulk_boundary_partition}) confirms Eq.(\ref{eq:bulk_boundary_partition}) as the appropriate AdS/CFT partition function to compute bulk correlators with boundary deformations.

To this end, given the construction of the bulk $\phi$ in the previous section as a lift of the boundary fields, we should equivalently be able to compute bulk correlators by using Eq.(\ref{eq:homogenous_bulk}) and computing the resulting correlators of $\alpha$ and $\beta$ using the boundary path integral in Eq.(\ref{eq:bulk_boundary_partition}).  This is reminiscent of the effort to construct bulk observables from boundary operators using smearing functions.  Before developing our lift formalism, it is worthwhile to first very briefly review the program and status of smearing functions.

\subsection{Smearing}

The goal of the smearing program is to construct bulk operators from their CFT duals through a \textit{linear} integral operation:
\begin{align}
\phi(B) = \int db K(B;b) \mO(b),
\end{align}
where the integral kernel $K(B;b)$ is the smearing function that integrates over a boundary coordinate $b$ to generate a field at bulk coordinate $B$.

Without interactions, $K(B;b)$ was found in global coordinates in \cite{Hamilton:2005ju} through both a Green function approach and mode function expansion.  A review of the mode expansion approach is give in \cite{Leichenauer:2013kaa} and is sketched here. 

A free bulk field may be generically expanded as
\begin{align}
\phi(B) = \sum_n f_n(B) a_n + h.c. \label{eq:phi_expansion}
\end{align}
where $f_n$ denotes the wave function (eigenfunction) with quantum numbers (eigenvalues) $n$ satisfying the classical bulk equations of motion, and $a_n$ ($a^\dagger_n$) is the associated annihilation (creation) operator.  With an appropriate normalization, $\{f_n\}$ forms an orthonormal set and $a_n$ consequently satisfies the appropriate algebra, $[a_n, a^\dagger_m] = \delta_{nm}$.

Carrying $\phi(B)$ to the boundary, $B\to b$, maps to $\mO(b)$ in the usual way, implying $\mO$ then inherits a related mode expansion:
\begin{align}
\mO(b) = \sum_n f_{\partial, n}(b) a_n + h.c.
\end{align}
Provided an appropriate foliation of AdS, $f_n$ can be defined to be orthonormal along radial slices on AdS and thus remain orthonormal at the boundary, implying $\langle f_{\partial, n} | f^\dagger_{\partial,m}\rangle = \delta_{n m}$.  With this, Eq.(\ref{eq:phi_expansion}) becomes
\begin{align}
\phi(B) = \sum_n  f_n(B) \langle f^\dagger_{\partial, n} \mO \rangle + h.c.
\end{align}
From this, the schematic form of the smearing function can be immediately extracted:
\begin{align}
K(B;b) = \sum_n f_n(B) f^\dagger_{\partial, n}(b) + h.c. \label{eq:smearing_scheme}
\end{align}
From Eq.(\ref{eq:smearing_scheme}), it immediately follows that $K(B;b)$ satisfies the classical equations of motion in the bulk.

The existence of $K(B;b)$ is predicated on the convergence and support of Eq.(\ref{eq:smearing_scheme}) for nontrivial $B$.  For certain backgrounds, such as in the presence of a black hole, the sum is non-convergent or lacks support \cite{Leichenauer:2013kaa}, and certain constructions result in a non-causal map ($\lim_{B\to b} K(B;b') \neq \delta(b-b')$) \cite{Bena:1999jv}. Nonetheless, smearing provides a powerful means of probing conformal theories dual to free AdS theories.  We aim to develop an alternative construction of bulk fields in the same spirit as smearing.



\subsection{Lifting}\label{sec:lift}

Instead of seeking a linear operation that maps $\mO$ to $\phi$, we employ Eq.(\ref{eq:homogenous_bulk}) with a caveat on the form of $\beta$ to be lifted that will be derived here.  To develop this formalism, consider first the undeformed bulk-boundary two point function:
\begin{align}
\epsilon^{-\Delta_-} \langle \phi(y, \epsilon) \phi(x, z) \rangle =& \int d^d x' \, \left[ L_\alpha (x, x'; z) \langle \alpha(y) \alpha(x') \rangle + 2\nu L_\beta (x, x'; z) \langle \alpha(y) \beta(x') \rangle \right].
\end{align}
The first boundary correlator is $\langle \alpha(y) \alpha(x') \rangle = g(y - x')$ (the boundary propagator); the second correlator must, perhaps surprisingly, evaluate to a local term, $\langle \alpha(y) \beta(x') \rangle = \frac{1}{2 \nu} \delta^d ( y - x')$, to obtain the correct bulk-boundary propagator.  However, integrating $\beta$ out in the absence of a deformation in Eq.(\ref{eq:bulk_boundary_partition}) sets $\beta=0$.  This suggests that, at the level of operators, we must make the substitution
\begin{align}
\beta \to \beta + \beta_0
\end{align}
in Eq.(\ref{eq:homogenous_bulk}). The action of this undeformed $\beta(x) $ on $f[\alpha]$ can be viewed as the functional derivative when evaluating correlators:
\begin{align}
\beta_0 (x) = \frac{1}{2\nu} \frac{\delta}{\delta \alpha(x)}.
\end{align}

Next, we evaluate the undeformed bulk-bulk two-point function using the same technique to find
\begin{align}
\langle \phi(-p, z) \phi(p, z') \rangle =& -(z z')^{d/2} \left[ I_{-\nu} ( p z) K_\nu (p z') + I_{-\nu} ( p z') K_\nu (p z) + I_{-\nu} ( p z) I_{-\nu} (p z') \right] \nonumber \\
&+ (2 \nu)^2 L_\beta . \langle  (\beta + \beta_0)  (\beta + \beta_0) \rangle . L_\beta ,
\end{align}
where the `$.$' binary operator denotes the integration over the common boundary coordinates of the adjacent objects. Since there are no interactions, the second line vanishes. The first of the three remaining terms is the two-point function for $z'>z$, the second is for $z>z'$, and the third is for when $z=z'$.  We have encountered a problem with treating the bulk as the lift of a CFT: the boundary has no knowledge of the relative $z-$positions of our correlators in the bulk.  This issue can be fixed by inserting a $z-$ordering operator, $Z$, into correlators: $\langle \phi \phi \rangle \to \langle Z \phi \phi \rangle$.  The operator annihilates $\beta_0$ for the field $\phi$ with the \textit{smaller} $z$ in Wick-contracted $\phi \phi$ pairs. It should be noted that the operator does only affects the non-interacting Witten diagrams as the diagrams contains boundary interactions are tautologically ordered.

With boundary deformations turned on, $\beta$ takes its usual form:
\begin{equation}
\beta = \frac{1}{2 \nu} W'[\alpha].
\end{equation}
With this feature, we find
\begin{align}
\langle Z \phi(x, z) \phi(x', z') \rangle =& G(x-x';z,z') + L_\alpha . \langle \alpha \alpha - \alpha_0 \alpha_0 \rangle . L_\alpha  + L_\alpha . \langle \alpha W'[\alpha] \rangle . L_\beta +  L_\beta . \langle W'[\alpha] \alpha \rangle . L_\alpha \nonumber \\
&+ L_\beta . \langle W'[\alpha](y) W'[\alpha](y') + W''[\alpha] \delta^d (y-y') \rangle . L_\beta, \label{eq:lifted_two_point}
\end{align}
where $\alpha_0$ is the undeformed $\alpha$.  It follows from pulling Eq.(\ref{eq:two_point_witten}) to the boundary, identifying the near-boundary bulk-boundary two-point function with
\begin{align}
\epsilon^{-\Delta_-} \langle \alpha(x) \phi(x', \epsilon) \rangle \approx& \langle \alpha(x) (\alpha(x') + \beta(x') \epsilon^{2\nu}) \rangle,
\end{align}
and demanding consistency among already confirmed results that, in momentum space,
\begin{align}
\langle \alpha W'[\alpha] \rangle = \frac{g \Sigma}{1 - g \Sigma}. \label{eq:alpha_beta}
\end{align}
With this identification, Eq.(\ref{eq:lifted_two_point}) reproduces Eq.(\ref{eq:two_point_witten}):
\begin{align}
\langle Z \phi(x, z) \phi(x', z') \rangle =& G(x-x';z,z') + \mathcal{K}(x-y;z) . \langle W'[\alpha](y) W'[\alpha](y') + W''[\alpha] \delta^d (y-y') \rangle . \mathcal{K}(y'-x';z'). \label{eq:lifted_two_point2}
\end{align}
All rules for computing bulk correlators from boundary data are now in place, completing the formalism.

It is worthwhile noting the subtle distinctions between lifting and smearing. While the lift and smear kernels both satisfy the linear classical bulk equations of motion in the absence of bulk interactions, lifting generally provides a nonlinear map from CFT operators to bulk fields in the presence of boundary interactions as a consequence of the boundary conditions and an affine map\footnote{A linear map is one of the form $y=mx$, while an affine map takes the form $y=mx+b$.} in the absence of deformations. The mapping also provides a means of reproducing the results of Witten diagrams at the level of correlators without the need to adjust a tower of coefficients to cancel noncausal effects in the presence of (boundary) interactions \cite{Kabat:2011rz}. The lift kernel also manifestly approaches a delta function when taken to the boundary, ensuring bulk fields constructed from boundary operators map correctly when taken to the boundary.

It is also worthwhile to recapitulate the results of this and the last few sections:
\begin{itemize}
\item The computation of bulk correlators in the presence of multi-trace boundary deformations should be carried out with Eq.(\ref{eq:bulk_boundary_partition}) as the generating partition function.

\item Pulling bulk correlators to the boundary in the usual way returns the same results as computing CFT correlators with conformal deformations.  That is to say, $Z_{CFT}[\phi_b] = \lim_{\epsilon \to 0} Z_{AdS}[\phi_b, \epsilon]$.

\item Bulk observables can be constructed from CFT observables via the lift provided by Eq.(\ref{eq:homogenous_bulk}).  The formalism makes the identification $\beta = \frac{1}{2\nu} \left[ W'[\alpha] + \frac{\delta}{\delta \alpha} \right]$ when computing correlators, and the operator $Z$ was introduced to order the lift operation by $z$-coordinate.

\item There is a strong connection between the $\beta$ term in the bulk and the 1PI diagram when computing correlators.  In particular, the results Eq.(\ref{eq:1PI_W}) and Eq.(\ref{eq:alpha_beta}) are interesting and useful.
\end{itemize}

\section{Dilatation spectrum and RG flow}\label{sec:rg_flow}

We now wish to use the results of the lift formalism to find a generic form for the conformal dimension of a dual operator $\mO$ as a function of energy scale in the presence of multi-trace deformations.  The results are the first steps to the multi-trace generalization of \cite{Fan:2011wm}.

In the absence of deformations, the dilatation spectrum is dual to the bulk mass spectrum via the mapping $m^2 = \Delta (\Delta - d)$.  The inclusion of conformal deformations triggers RG flow such that the IR spectrum can often be extracted from the UV spectrum.  For instance, in the presence of double-trace deformations, operators with dimension $\Delta_-( = \frac{d}{2} - \nu \le \frac{d}{2})$ in the UV flow to a $\Delta_+( = \frac{d}{2} + \nu)$ fixed point in the IR \cite{Gubser:2002vv}.  The conformal dominance program exploits the UV conformal basis to construct IR mass states for certain deformations \cite{Katz:2013qua,Hogervorst:2014rta}.  In what follows we restrict the scaling dimension to $\Delta = \Delta_-$.  Many results can be immediately extended to $\Delta=\Delta_+$, however we are chiefly concerned with the RG flow of the CFTs between potential fixed points.

Dilatation eigenstates in the undeformed CFT are created by placing an operator at the origin: $| 0 \rangle = \mO(0) |\Omega \rangle$. It follows from the identity $[ D, \mO(x) ] = \left( x \cdot \partial + \Delta_- \right) \mO(x)$ that
\begin{align}
\langle x | D | 0 \rangle = \Delta_- \langle x | 0 \rangle = -\left( x \cdot \partial + \Delta_- \right) \langle x| 0 \rangle.
\end{align}
Evidently, the CFT two-point functions are eigenfunctions of the differential representation of the dilatation operator.

When lifting to the bulk, the scaling dimension is replaced by differentiation with respect to $z$: $(x \cdot \partial + \Delta_-)\mO \to (x \cdot \partial + z \partial_z) \phi$.  With this replacement, the bulk-boundary propagator is found to be an eigenfunction of the dilatation operator:
\begin{align}
\langle \phi(x,z)| D | 0 \rangle = - \left( x \cdot \partial + z \partial_z \right) \mathcal{K}(x;z) = \Delta_- \mathcal{K}(x; z).
\end{align}
Even more readily, and perhaps crucially, the classical field $\phi$ is an eigenfunction of $z \partial_z$ as $z \to 0$ with eigenvalue $\Delta_-$.

When deformations are introduced, the story becomes more complex.  In momentum space, the boundary two-point function remains an eigenfunction of the dilatation operator, but the eigenvalue gains a momentum-dependence:
\begin{align}
\left( p \cdot \frac{\partial}{\partial p} + \Delta_- \right) \langle p | 0 \rangle = \left[ \Delta_- + \frac{g(p)}{1 - g(p) \Sigma(p)} \left( p \cdot \frac{\partial}{\partial p} - 2 \nu\right) \Sigma(p) \right] \langle p | 0 \rangle. \label{eq:boundary_conformal_dimension}
\end{align} 
As expected for a double trace deformation ($\Sigma = const.$) in the UV, we find $g \Sigma \ll 1$, which leads to $\Delta = \Delta_-$; in the IR, we find$g \Sigma \gg 1$, which leads to $\Delta = \Delta_+$.

The bulk-boundary two point function also remains an eigenfunction of the dilatation operator with the same eigenvalue as its boundary counterpart:
\begin{align}
&\left( p \cdot \frac{\partial}{\partial p} + z \partial_z \right) \frac{1}{1 - g(p) \Sigma(p)} \mathcal{K}(p;z) = \nonumber \\
&\left[ \Delta_- + \frac{g(p)}{1 - g(p) \Sigma(p)} \left( p \cdot \frac{\partial}{\partial p} - 2 \nu\right) \Sigma(p) \right]  \frac{1}{1 - g(p) \Sigma(p)} \mathcal{K}(p;z).\label{eq:ads_delta}
\end{align}
The relation between the bulk operator $z \partial_z$ and the conformal dimension becomes more obscure in the presence of deformations.  We could attempt to demand the field $\phi$ remain the eigenfunction of the operator at the boundary designated by the $z=\epsilon$ cutoff and identify the eigenvalue with $\Delta$.  Carrying through with the procedure with a double-trace deformation and keeping next to leading order terms in $\epsilon$ for $\phi$ results in
\begin{align}
\Delta = \phi^{-1} \epsilon \partial_\epsilon \phi = \Delta_- + \frac{\lambda \epsilon^{2\nu}}{1 + \frac{\lambda}{2\nu}  \epsilon^{2\nu}}. \label{eq:rg_double_trace}
\end{align}
If we interpret the UV brane on which the bulk theory terminates as the inverse of the renormalization scale, $\epsilon \propto \mu^{-1}$, the RG flow in Eq.(\ref{eq:rg_double_trace}) matches exactly Eq.(\ref{eq:boundary_conformal_dimension}) for double trace deformations.  The procedure as presented is serendipitous for double-trace deformations since $\beta$ is classically linear in $\alpha$, allowing the field to completely divide out; this does not occur for more complicated deformations.  However, the promising connection between the $z$-direction in the bulk and RG flow at the boundary begs for the procedure to be generalized.

When transitioning to the quantum formulation, we should expect to deal with correlators of the fields instead of the fields themselves. Classically, we may multiply and divide Eq.(\ref{eq:rg_double_trace}) by $\alpha$, so transitioning suggests we compute $\Delta$ in momentum space by pulling the bulk-boundary propagator to the UV cutoff and evaluating the momentum at this UV scale: 
\begin{align}
\Delta   \underset{\epsilon \to 0}{=}& \left. \frac{\epsilon \partial_\epsilon \left[ \langle \phi \alpha \rangle (p, \epsilon) \right]}{ \left[ \langle \phi \alpha \rangle (p, \epsilon) \right]} \right|_{|p| = \epsilon^{-1}} \label{eq:ads_rg} 
\end{align}
for $\epsilon \to 0$. This procedure actually trivially follows from Eq.(\ref{eq:ads_delta}) by simply demanding the momentum be evaluated at the UV cutoff. Physically, Eq.(\ref{eq:ads_rg}) says that the scaling dimension is a measure of how the bulk-boundary propagator changes as the location of the UV brane is shifted while keeping the energy near the renormalization scale.

Using Eqs.(\ref{eq:homogenous_bulk})\&(\ref{eq:beta_redux}) to expand the bulk fields in Eq.(\ref{eq:ads_rg}) in terms of $\alpha$, inserting Eq.(\ref{eq:alpha_beta}) into the result, and setting $\epsilon \to \mu^{-1}$ yields
\begin{align}
\Delta =& \Delta_- + \frac{  \frac{\mu^{-2\nu}}{2 \nu} \left[ 2\nu - \mu \partial_\mu \right] \Sigma(\mu)}{1 + \frac{\mu^{-2\nu}}{2\nu} \Sigma(\mu)}. \label{eq:deformed_delta2}
\end{align}

Once again, the correct RG flow has been recovered.  This indicates that the running of a deformed CFT from the UV to the IR may be studied in AdS by ending the theory on a brane at $z=\epsilon$ and computing appropriate quantities by setting the renormalization scale $\mu\to \epsilon^{-1}$. This procedure is in independent agreement with the holographic RG program of interpreting the classical evolution of fields in the radial direction in AdS as RG flow at the boundary.  The dilatation spectrum can be explicitly computed in the bulk using Eq.(\ref{eq:ads_rg}), and we wish to emphasize the necessity of the $\beta_0$ piece defined via the lift formalism in computing the bulk correlators.  It is not difficult to imagine a similar analytical procedure should hold for the mass-squared spectrum, although the details are not immediately apparent.

\section{Discussion}

We have constructed rules utilizing modified boundary conditions in AdS to compute bulk correlators through the use of an appropriate AdS/CFT partition function and via a lift formalism for theories subject to a conformal deformation.  The partition function explicitly relates the AdS theory to the CFT theory and establishes the AdS side in the absence of bulk interactions as a theory that evolves classically in the $z$- (radial-) direction with boundary conditions subject to quantum effects.  The lift formalism, as an alternative to smearing, provides the inverse of the usual boundary-scaling dictionary that relates the bulk and boundary operators ($\mO = \lim_{z \to 0} z^{-\Delta} \phi(z)$).  Utilizing the results of the lift formalism, a formula to compute the conformal dimension of CFT operators as a function of energy scale using AdS correlators was derived.

We have not considered the obstructions that may hinder obtaining an appropriate smearing kernel.  While we expect the lift formalism to fail or require modification when bulk interactions are turned on as the bulk would no longer evolve classically in the z-direction, it is our hope that it remains valid for asymptotically AdS spaces so that semiclassical gravity may be studied.

For now, only the running of the scaling dimension with the renormalization scale was considered since its fundamental role in the AdS/CFT correspondence makes it easy to handle.  A similar strategy of finding an appropriate differential operator in the bulk and writing down a ratio of correlators may likely be employed to compute the dependence of mass-squared elements on the renormalization scale to approach conformal dominance from a bulk perspective.  It would be of interest to explore this approach in the context of the c-theorem.



\section{Acknowledgments}
The author would like to thank Christopher Brust, Nikhil Anand, and Jared Kaplan for useful discussions. The author was partially supported by NSF grant PHY-1316665.

\appendix
\section{Classical fields with multi-trace deformations}\label{app:classical}

We can extend the formalism of \S \ref{sec:classical} to determine the bulk-boundary propagator with multi-trace boundary deformations. Consider the following boundary deformation and associated boundary conditions:
\begin{align}
\mathcal{L}_b = \epsilon^{d-\Delta + 1} \phi_b \phi + \frac{1}{n} \lambda \epsilon^{d - n \Delta + 1} \phi^n \implies \delta B[\phi] = - \epsilon^{d - \Delta} \phi_b - \lambda \epsilon^{d - \Delta}  \alpha^{n-1}.
\end{align}
The usual procedure results in
\begin{align}
b(x) = \frac{1}{2\nu} \left[ \phi_b(x) + \lambda \alpha^{n-1}(x) \right],
\end{align}
and we face solving
\begin{align}
\alpha(x) =& \int d^d x' \,\int \dbar^d p \, \frac{\Gamma(\nu)}{\Gamma(1-\nu)} \left( \frac{p}{2} \right)^{-2 \nu} \frac{1}{2} \left[ \phi_b(x') + \lambda \alpha^{n-1}(x') \right] e^{i p \cdot ( x - x')} \label{eq:npoint_solution1}
\end{align}
for $\alpha$ in terms of $\phi_b$.  Determining $\alpha$ exactly seems like a hopeless endeavor, but we can still compute the contribution of the deformation to the bulk-boundary propagator. Given the nonlinearity we should expect these deformations to generate interaction terms in higher-point correlation functions.  Classically, the interacting piece of $N$-point functions can be computed as contributions to the $N$th harmonic of the source.  This follows explicitly from
\begin{align}
\langle \alpha(x_1) \dots \alpha(x_N) \rangle \propto \left.  \frac{\delta}{\delta \phi_b(x_1)} \dots \frac{\delta}{\delta \phi_b(x_N)} e^{ \int d^d y\,  \alpha(y) \phi_b(y)} \right|_{\phi_b =0}, \label{eq:npointfunc}
\end{align}
the purely interacting piece of which is
\begin{align}
\left. \frac{\delta}{\delta \phi_b(x_2)} \dots \frac{\delta}{\delta \phi_b(x_N)} \alpha(x_1) + \dots + \frac{\delta}{\delta \phi_b(x_1)} \dots \frac{\delta}{\delta \phi_b(x_{N-1})} \alpha(x_N) \right|_{\phi_b=0}. \label{eq:npointfunc_int}
\end{align}
We can build an expansion of Eq.(\ref{eq:npoint_solution1}) in nested functionals of $\phi_b$, which poises us perfectly to compute $N$-point functions as sums of products of integral kernels in accordance with Eqs.(\ref{eq:npointfunc})\&(\ref{eq:npointfunc_int}).  The bulk-boundary propagator would then simply be the first functional derivative of $\phi$ with respect to $\phi_b$. Since $\alpha[\phi_b=0]=0$, there are no classical contributions to the bulk-boundary propagator for $n>2$. 

Computing higher point correlation functions in AdS generated by boundary deformations, of course, require us to consider bulk sources. We can expect to use similar methods to determine these.  Heuristically, even classically an $N$-point function in the bulk should involve an $N$-point function at the boundary, and, from Eq.(\ref{eq:npoint_solution1}), we expect a non-vanishing tree-level correlator only for $N=n$.  There is more in the details, but at the level of the bulk-boundary propagator, we can stop here.

In the absence of boundary deformations, the wave functions are well known \cite{Sundrum:2011ic,Anand:2015zea}:
\begin{align}
f^0( \vec{p}, m; x, z) =& \sqrt{\frac{i m}{2 p_{m 0}}} z^{\frac{d}{2}} J_{-\nu}(m z) e^{i p_m \cdot x},
\end{align}
where $p_{m 0} = i \sqrt{ \vec{p}^2 + m^2}$.  As was the case with the bulk-boundary propagator, we can use this knowledge to compute the modified wave functions in the presence of boundary deformations.

The procedure works as before with $\phi \to f$, except we exclude boundary source terms from our Lagrangian and boundary conditions,
\begin{align}
\mathcal{L}_b =  \frac{1}{n} \lambda \epsilon^{d - n \Delta + 1} \phi^n \implies \delta B[\phi] = - \lambda \epsilon^{d - \Delta}  \alpha^{n-1},
\end{align}
and we seek homogeneous solutions. Now, Eq.(\ref{eq:npoint_solution1}) becomes
\begin{align}
&\alpha(x) = \sqrt{\frac{i m}{2 p_{m 0}}} \frac{1}{\Gamma( 1 - \nu )} \left( \frac{m}{2} \right)^{- \nu} e^{i p_m \cdot x} +  \int d^d x' \,\int \dbar^d p \, \frac{\Gamma(\nu)}{\Gamma(1-\nu)} \left( \frac{p}{2} \right)^{-2 \nu} \frac{\lambda }{2}  \alpha^{n-1}(x') e^{i p \cdot ( x - x')} \label{eq:wave_solution1}.
\end{align}
For $n=2$, this is straight forward:
\begin{align}
\alpha(x) =& \frac{1}{1 - \frac{\Gamma(\nu)}{\Gamma(1-\nu)} \left( \frac{p_m}{2} \right)^{-2\nu} \frac{\lambda}{2}} \sqrt{\frac{i m}{2 p_{m 0}}} \frac{1}{ \Gamma( 1 - \nu )} \left( \frac{m}{2} \right)^{- \nu} e^{i p_m \cdot x}.
\end{align}
And so
\begin{align}
f(x,z) =& \sqrt{\frac{i m}{2 p_{m 0}}} z^{\frac{d}{2}} \Bigg[J_{-\nu}(m z) \nonumber \\
&+ \frac{\lambda}{1 - \frac{\Gamma(\nu)}{\Gamma(1-\nu)} \left( \frac{p_m}{2} \right)^{-2\nu} \frac{\lambda}{2}}  \frac{1}{ \Gamma( 1 - \nu )^2} \left( \frac{m}{2} \right)^{- \nu} \left( \frac{p_m}{2} \right)^{-\nu}   K_\nu(p_m z) \Bigg]e^{i p_m \cdot x}.
\end{align}

It is worthwhile to point out that Eq.(\ref{eq:wave_solution1}) requires $\alpha$ to be classical at the boundary, demonstrating that the classical methods sufficient for double-trace deformations are inapplicable for more general deformations.

\bibliographystyle{utphys}
\bibliography{biblio}

\end{document}